\documentstyle[prb,aps,graphicx,floats]{revtex}

\begin{document}
\draft

\title{
Stable ultrahigh-density magneto-optical recordings using 
introduced linear defects}

\author{L. Krusin-Elbaum$^{\star\dag}$, T. Shibauchi$^{\star \dag\S}$, B. Argyle$^{\star}$, L. Gignac$^{\star}$
\& D. Weller$^{\ddag\parallel}$}

\address{$^{\star}$IBM T.~J. Watson
Research Center, Yorktown Heights, NY 10598, USA}
\address{$^{\S}$MST-STC, Los Alamos National Laboratory, MS-K763, Los Alamos, NM 87545,
USA}
\address{$^{\ddag}$IBM Almaden Research Center, San Jose, CA 95120,
USA}

\address{$^{\dag}$These authors contributed equally to this work}


\wideabs{ {\rm \begin{flushright}Nature {\bf 410}, 444 (March 22, 
2001).\end{flushright}} 

\maketitle } 

\narrowtext

\vspace{6mm}

\noindent {\bf The stability of data bits in magnetic recording 
media \cite{daniel,thompson} at ultrahigh densities is 
compromised by thermal `flips' -- magnetic spin reversals -- of 
nano-sized spin domains \cite{Hubert}, which erase the stored 
information. Media that are magnetized perpendicular to the plane 
of the film, such as ultrathin cobalt films or multilayered 
structures \cite{kirby,allenspach1}, are more stable against 
thermal self-erasure \cite{thompson,weller99} than conventional 
memory devices. In this context, magneto-optical memories seem 
particularly promising for ultrahigh-density recording on 
portable disks, and bit densities of $\sim$100 Gbit inch$^{-2}$ 
(ref. \onlinecite{partovi99a}) have been demonstrated using 
recent advances in the bit writing and reading techniques 
\cite{partovi99a,awano,betzig91,partovi99,Kaneko}. But the 
roughness and mobility of the magnetic domain walls 
\cite{jaggedness,lemerle} prevents closer packing of the magnetic 
bits, and therefore presents a challenge to reaching even higher 
bit densities. Here we report that the strain imposed by a linear 
defect in a magnetic thin film can smooth rough domain walls over 
regions hundreds of micrometers in size, and halt their motion. A 
scaling analysis of this process, based on the generic physics of 
disorder-controlled elastic lines 
\cite{healy,hwa,Blatter,krusin1}, points to a simple way by which 
magnetic media might be prepared that can store data at densities 
in excess of 1 Tbit inch$^{-2}$.}

Increasing information storage to densities past 100 Gbit 
inch$^{-2}$ may evolve through extensions of current magnetic 
recording technologies (to patterned media \cite{weller99}, for 
example). But such increases in storage density might be achieved 
by using other techniques such as holography (via interference 
patterns produced by two intersecting laser beams) 
\cite{holography}, or micromachined nano-cantilever arrays 
\cite{binnig}, or -- to satisfy a relentless demand for 
portability -- using scanning localized subwavelength ($< 
\lambda/40$) (near-field) optical probes \cite{betzig91} that can 
imprint and resolve images in magneto-optic media \cite{awano} of 
the order of the probe size. The bit-writing with local probes 
may be thermally assisted by a current \cite{nakamura} or a laser 
beam that raises local temperature to the vicinity of the Curie 
temperature $T_C$, resulting in the formation of a reversed 
domain with a rough wall. To realize smooth and (for precise 
positioning) stable (`noiseless') domain walls that can be 
implemented without, for example, nanoscale patterning, we 
suggest a conceptual identification of walls with elastic lines 
\cite{Blatter} and seek to utilize linear defects inducing a 
directional long-range strain field in ultrathin Pt/Co/Pt 
structures -- long considered excellent candidates for 
high-density recording at blue-range wavelengths \cite{Kaneko}. 
In such trilayers (Fig. 1), large uniaxial perpendicular 
anisotropy is sustained by the interface contribution $\propto 
1/d_{\rm Co}$ up to Co thickness $d_{\rm Co} \sim$ 1.5~nm; beyond 
this thickness the magnetizaton switches from out-of-plane to 
in-plane \cite{allenspach1}.

Figure 2a shows two typical up-domains (see legend for imaging  
details \cite{bernie}), nucleated owing to a locally suppressed 
coercive field \cite{Hubert} $H_c$ in a 0.7~nm Co film sandwiched 
between 3~nm (top) and 2~nm (bottom) Pt layers, prepared using 
standard deposition parameters \cite{Weller} and without linear 
defects. The domains are round with (undirected) domain walls 
(DWs) that are as rugged as expected \cite{jaggedness}. We note 
that patterned nucleation sites, as in Fig. 2b-d, do not reduce 
either DW roughness or velocity. As magnetic field is increased 
beyond $H_c = 750$~Oe (inset in Fig. 3), the outward motion of 
DWs becomes increasingly swift (estimated from displacement of a 
small segment of the wall during 500-ms field pulses, the wall 
velocity is more than $180~ \mu$m s$^{-1}$ at $H = 854$~Oe).

To control the rugged DW structure we will use a recently found 
connection between the DW behavior in thin films \cite{lemerle} 
and that of elastic lines \cite{healy,hwa,Blatter}. One prominent 
example of directed elastic lines are wandering vortex lines 
(quantized magnetic flux lines maintained by a swirling tube of 
current) in high temperature superconductors \cite{Blatter} that 
can be strongly localized by interaction with columnar (linear) 
defects \cite{krusin1}. A powerful arsenal of ideas can now be 
engaged to understand how a linear defect potential may localize 
and reduce the roughness of DWs, and force them to accommodate to 
the defect shape. We introduce a line defect that delivers a 
three-punch action. Through magneto-elastic coupling, (i) it 
gives the wall a preferred direction \cite{healy} (along the 
defect), (ii) it increases its elasticity, reducing the DW 
roughness as it negotiates the random landscape, and (iii) it 
acts to reduce wall velocity to a nearly full stop in fields 
greater than the coercive field of unmodified film. The defect is 
installed during the Co deposition by imposing an anisotropic 
tension (clamping) on the substrate and its subsequent release 
\cite{stress}. The resulting Co `fold' (Fig. 2g,h) introduces a 
$y$-axis invariant long-range strain field $\varepsilon(x)$.

A DW driven by the magnetic field toward such linear defect (Fig. 
2e,f) presents a structural contrast with the DWs in the 
unmodified film in Fig. 2a. Even at large distances ($\sim 
300~\mu$m) away from the defect, the DW conforms on the average 
to the defect line along $y$. It becomes progressively smoother 
(and straighter) as it approaches the defect. It also rapidly 
decelerates. The deceleration and near-standstill of the wall 
depends on the proximity to the defect, as represented by the 
spatial progression of the velocity-versus-field (force) response 
in Fig. 3. The $v$-$H$ curves are highly nonlinear, as has been 
reported recently in Co films \cite{lemerle} where the disorder 
landscape is formed, for example, by atomic scale imperfections 
at the film--substrate interface. Such nonlinear response in the 
limit of vanishing driving force is a signature of glassy (creep) 
dynamics \cite{vinokur}, well established for the elastic vortex 
lines in a superconductor through measurements of the 
voltage-versus-current ($V$-$I$) characteristics \cite{Blatter}. 
Above $H_{crit}$ (obtained from the $v$-$H$ curves by the usual 
velocity cut-off criterion, here chosen at $v = 0.14~\mu$m 
s$^{-1}$) the driving field exceeds the `pinning' force 
\cite{Blatter} and the DW response becomes linear and faster. We 
note that DW velocity, even far away from a line defect, is 
orders of magnitude lower than in unmodified films. Figure 3 
shows a field-forward advance of $v$-$H$ curves to higher fields, 
and an enhanced $H_{crit}$ (often referred to as a `propagation 
field' \cite{Hubert}) on the approach to the line defect. An 
effective potential well that localizes the wall is formed by the 
driving field pushing the wall and the line-defect that acts 
against this push. It resembles columnar defects in a 
superconductor, where the critical current $J_{crit}$ is enhanced 
\cite{krusin1}. $H_{crit}$ correlates with the long-range 
repelling force field $H(x)$ exerted on the DWs by the 
line-defect, whose spatial extent is mapped in Fig. 4a. The shape 
of $H(x)$ -- a ridge along $y$ -- is either extracted directly 
from Kerr images taken with increasing $H$ after a fixed 
propagation time $t$ (main panel), or more quantitatively from 
the averaged DW positions ${\langle x\rangle}_y$ versus time at 
all fields. As illustrated in the inset, it takes a higher field 
$H$ to get closer to the line-defect; but at any $H$, after $t 
=1,800$~s, two DWs -- one approaching the ridge from the left and 
another from the right -- become effectively stationary.

Within the elastic description of a DW, the relevant scale is a 
collective pinning \cite{vinokur} length 
$L_c=(\epsilon_{el}^2\xi^2/\Delta)^{1/3}$, where $\xi$ and 
$\Delta$ are characteristic size and strength of underlying 
random disorder, and $\epsilon_{el}$ is the wall energy. At 
lengths $L>L_c$, the wall will elastically adjust to the random 
landscape to nestle in a local minimum energy configuration. The 
DW energy density (in addition to a uniform field term and 
ignoring a weak dipolar term) can be written as a sum of three: 
the exchange energy $\gamma_{ex}(x)$, the anisotopy energy 
$\gamma_{an}(x)$, and the magneto-elastic energy 
\cite{anisotropy} $\gamma_{m{\rm -}el}(x)$ coupling to the strain 
$\varepsilon(x)$ generated by the line-defect,

\begin{equation}
\gamma_{DW} = ~{\underbrace {A{(d\theta/dx)}^2}_{\gamma_{ex}}}~~~ 
{\underbrace {-K'm_z^2}_{\gamma_{an}}}~~~ {\underbrace 
{-B\varepsilon(x)m_z^2}_{\gamma_{m{\rm -}el}}}~~+H^2/8\pi. 
\label{DWenergy}
\end{equation}

Here $A$ is the exchange stiffness \cite{Hubert}, $K' = K - 2\pi 
M_S^2$ ($K$ is the anisotropy constant and $M_S$ is 
Pt-polarization-enhanced saturation magnetization of Co), and $B$ 
is related to Young's modulus of the film. Magnetization ${\vec { 
M }}$ rotates from `up' to `down' within the wall thickness 
$\delta$ (we estimate at $\sim$3~nm) and the wall (in the 
simplest form) is of Bloch type \cite{lemerle}, where the 
azimuthal angle $\phi = 0^\circ$ and the rotation is 
parameterized by an angle $\theta$ between the $z$-axis and 
${\vec { M }}$ ($m_z$ is the direction cosine along $z$). 
Minimizing \( {\displaystyle \int\!\gamma_{DW}dx} \) leads to DW 
energy \cite{Hubert} ${\cal E}_{DW} = 4\sqrt{AK'_{eff}}$, where 
$K'_{eff} = K' +B\varepsilon(x)$. The total wall energy (per unit 
area) in a magnetic field $H$ is $~{\cal E} = {\cal 
E}_{DW}-2M_SHx$. From the stability condition $d{\cal E}/dx = 0$ 
we obtain a nonlinear differential equation for $\varepsilon(x)$: 
$ H(x) = A{M_S}^{-1}B(d\varepsilon/dx){{(A K'_{eff})}}^{-1/2}$, 
which we solve numerically using the stationary $H(x)$ mapped in 
Fig. 4a. The result is a factor of $\sim 2$ enhanced effective 
anisotropy $K'_{eff}$ (Fig. 4b) and, consequently, an enhanced DW 
elastic energy (per unit length) \cite{lemerle} 
$\epsilon_{el}=4\sqrt{AK'_{eff}}d_{\rm Co}$. Thus, the enhanced 
elasticity of DW is in part responsible for the longer $L_c$. The 
longer $L_c$ is a key to the reduced DW roughness. This can be 
seen from the analysis of transverse displacements of the domain 
wall segments by computing from the DW images a (line shape) 
correlator \cite{healy} $x_t = \sqrt{<[x(y)-x(y+L)]^2>}$, 
predicted for $L < L_c$ to scale as \cite{Blatter} $ x_t \propto 
{(L/L_c)}^{3/2}$. We find that the characteristic length $L_c$ is 
enhanced in the vicinity of the line-defect by a factor of $\sim 
5$ (Fig. 4c). This implies a reduction of roughness, as measured 
by $x_t$, by a factor of $\sim 10$ on the shortest length scales 
and hence a potential increase in the areal density by nearly two 
orders of magnitude. In its proximity, the line defect wins over 
randomness so that on sufficiently long distances along its 
length DWs become essentially `flat' (that is, finite $x_t$ for 
$L \rightarrow \infty $) \cite{healy} in fields up to $\sim 2 
H_c$. The wall-smoothing at higher fields and at speeds of 
current recording technology ($H \approx 3 H_c, ~\sim 1~$Gbit 
s$^{-1}$) deserves further study for practical implementations of 
this effect. The long-range strain associated with a line defect 
presents an efficient, and relatively simple and cost-effective 
control of domain walls, with only a few such defects needed to 
stabilize large areas of ultrathin films.

\vspace{10mm}

\noindent {\bf Acknowledgements}

\small

We thank J. Slonczewski, V. Vinokur, G. Blatter, and G. Zimanyi for useful discussions.
Correspondence and requests for materials should be addressed to L. Krusin-Elbaum, fax:
+1-914-945-2141; e-mail: krusin@us.ibm.com.

$^{\parallel}$ present address: Seagate Technology, Pittsburgh, PA 15203, USA.

\onecolumn 

\normalsize
\begin{figure}[tb]
\begin{center}
\includegraphics[width=5cm]{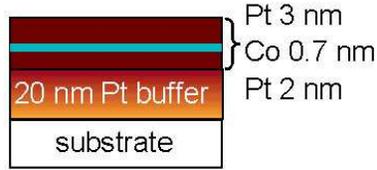}
\end{center} 
\caption{A sketch of an ultrathin Pt/Co/Pt trilayer stack 
explored in this study. Stacks were electron-beam evaporated at 
$190^\circ$C on glass or SiN$_x$/Si substrates with a 20~nm (111) 
textured Pt buffer layers, as described in ref.~22. }

\label{fig:1}
\end{figure}


\begin{figure}[tb]
\begin{center}
\includegraphics[width=18cm]{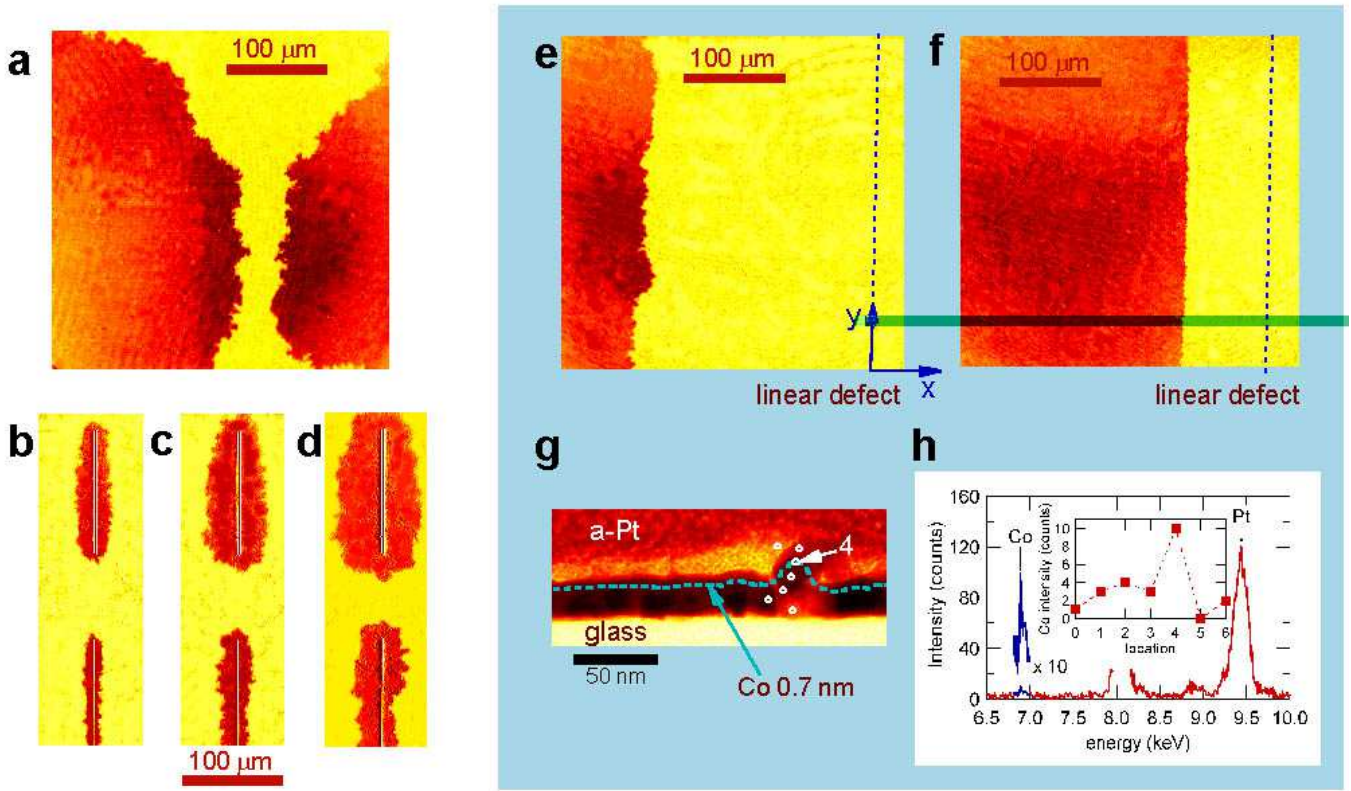}
\end{center}
\caption{Effect of a linear defect on domain walls in Pt/Co/Pt 
films. Polar magneto-optic Kerr microscopy 
\protect{\cite{Hubert}} images were obtained using a 3-W argon 
laser illumination ($\lambda$ = 514.5 $\&$ 488.0 nm) to enhance 
phase contrast with a laser-beam tandem-dithering technique 
\protect{\cite{bernie}} to eliminate the effects of laser 
coherence. We first saturated magnetization with a negative 
$z$-axis magnetic field and then quenched field to zero. After a 
first positive square field pulse, an up-domain was nucleated and 
its motion was imaged at short time intervals in fields up to 
$\sim 2~$kOe. {\bf a}, Typical room-temperature image of domains 
in a film without line defects after a 854-Oe magnetic field 
applied for 1 s was removed. The walls show roughness on length 
scales much larger than the grain size ($\sim 20$-$30~$nm) of the 
Pt buffer. The rapid motion of domain walls (DWs) is illustrated 
in snapshots in field $H = 616~$Oe after {\bf b} 1 s, {\bf c}, 4 
s and {\bf d}, 8 s of DWs that were seeded with elongated defects 
installed with a 30-keV Ga$^+$ FIB (Focused Ion Beam). DW 
roughness is independent of the seed-defect width (150~nm for 
bottom and 1~$\mu$m for top domains). {\bf e}, {\bf f},  A 
remarkable reduction in the domain wall roughness and speed is 
illustrated when a line-defect (dashed line along $y$ axis) is 
introduced. The images here were taken at $H = 924~$Oe ($>H_c = 
750~$Oe in {\bf a}] at {\bf e}, 40 s and {\bf f}, 1,800 s. The 
domain wall becomes directed along the defect. A cross-sectional 
electron transmission micrograph ({\bf g}) of an $\sim 
30$-nm-thin FIBed section -- normal to the line-defect -- shows 
an asymmetric `fold' of the trilayer stack. An amorphous Pt 
(a-Pt) cap was deposited to protect the trilayer during the FIB 
process. The $\sim 10$-nm elevation of Co in the `fold' is traced 
by the X-ray microprobe ({\bf h}) at spots such as are marked by 
six white circles in {\bf g} (the bottom is assigned `0'). Co 
intensity versus spot number is plotted in the inset of {\bf h}. }

\label{fig:2}
\end{figure}

\begin{figure}[tb]
\begin{center}
\vspace{-8mm}
\includegraphics[width=10cm]{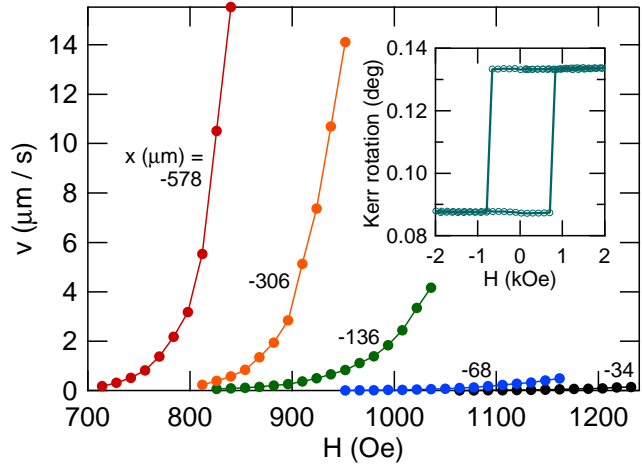}
\end{center}
\caption{Domain wall velocity as a function of applied magnetic 
field for different values of $x$. Far away ($-$578~$\mu$m) from 
the line-defect `ridge' at $x = 0$, the velocity takes off 
rapidly with field near $H_{crit}$. Well below $H_{crit}$, the 
nonlinear $v$-$H$ curves follow \cite{lemerle} $v(H) \propto 
\exp[-(U_c/k_BT){(H_{crit}/H)}^\mu]$, where $U_c$ is the 
collective pinning energy scale, $k_B$ is the Boltzmann constant, 
$T$ is temperature, and $\mu = 1/4$ is the glassy dynamical 
exponent related to the roughness exponent, see ref.~14. 
$H_{crit}$ grows on the approach to the line defect (see Fig. 
4a). Inset: A square Kerr rotation (magnetization) hysteresis 
characteristic of perpendicular anisotropy of our film in Fig. 
2a.}

\label{fig:3}
\end{figure}

\begin{figure}[tb]
\begin{center}
\vspace{-8mm}
\includegraphics[width=12cm]{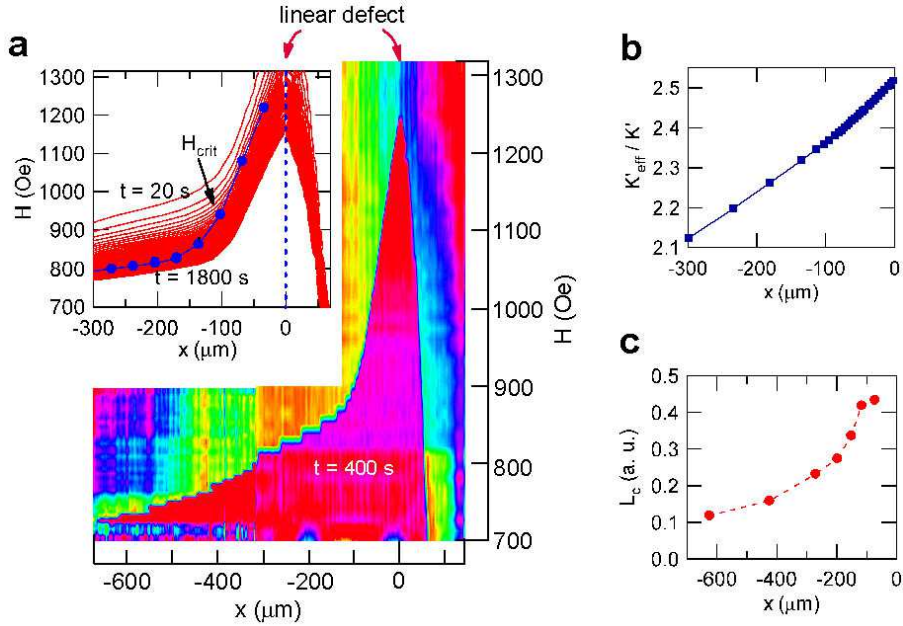}
\end{center}
\caption{A force field $H(x)$ generated by a line-defect along 
$y$ repels a magnetic-field-driven domain wall. {\bf a}, A fixed 
$y$-cut from a map of the defect ridge obtained from combining 
domain images for different fields after 400~s. To display the 
long range ($x = -680$ to $+140$~$\mu$m) of the force field of 
the line defect, we overlapped two series of Kerr observations 
($H = 700$ to $1,316~$Oe): one containing a line defect at $x=0$ 
(as in the inset) and another shifted to the left by 400~$\mu$m. 
The ridge is mapped by imaging two domains approaching it from 
both sides. Contours (at 20-s intervals) of the defect field 
(inset) that were obtained by extracting DW positions at 
different times and fields show how two DWs on two sides of the 
ridge become stationary when defect field balances the driving 
field, defining the ridge's outline. The propagation field 
$H_{crit}$ (solid blue dots), obtained from $v$-$H$ curves in 
Fig. 3 is enhanced by the line defect. The line defect 
effectively enhances ({\bf b}) the anisotropy constant $K'_{eff}$ 
and elongates ({\bf c}) the scaling length $L_c$. }

\label{fig:4}
\end{figure}

\end{document}